# ALIVE: A Low-Cost Interactive Vaccine Storage Environment Module ensuring easy portability and remote tracking of operational logistics to the last mile

Arkadeep Datta*, Arani Mukhopadhyay, Amitava Datta and Ranjan Ganguly

Advanced Materials Research and Applications (AMRA) Laboratory,
Department of Power Engineering, Jadavpur University (SL Campus), Kolkata – 700106
*Corresponding: arkadeepdatta@gmail.com

**Abstract.**
The COVID-19 pandemic has profoundly reshaped our lives, prompting a search for solutions to its far-reaching effects. Vaccines emerged as a beacon of hope, yet reaching remote areas faces last-mile hurdles and cost issues due to loss of vaccine potency due to poor temperature regulation of the storage units and unanticipated vaccine wastage *en route*, a common occurrence in conventional vaccine transportation methods. To ameliorate this problem, we introduce ALIVE, a low-cost Interactive Vaccine Storage Environment module. ALIVE provides an off-grid, self-sufficient solution for vaccine storage and transport, enabled by active cooling technology. ALIVE's innovation lies in its integration with the Internet of Things (IoT), allowing real-time monitoring and control. This IoT-enabled Application Programming Interface (API) features a data acquisition and environment parameter control system, managing oversight and decision-making. ALIVE's compact, lightweight design makes it adaptable to various logistical scenarios, while its versatility enables it to maintain both time-invariant and time-dependent thermophysical and spatial parameters. Operationalized through a PID algorithm, ALIVE ensures precise temperature control within the vaccine chamber. Its dynamic features, such as remote actuation and data sharing, demonstrate its adaptability and potential applications. Despite the frugal nature of development, the system promises significant benefits, including reduced vaccine loss and remote monitoring advantages. Collaborations with healthcare partners seek to further enhance ALIVE's readiness and expand its impact. ALIVE revolutionizes vaccine logistics, offering scalable, cost-effective solutions for bridging accessibility gaps in challenging distribution scenarios. Its adaptability positions it for widespread application, from last-mile vaccine delivery to environment-controlled supply chains and beyond.

## 1    Introduction

Amidst the dreaded spell of SARS-CoV-2 pandemic, commonly known as COVID-19, our lives underwent a profound transformation [1], impacting every facet of our daily routines. Amidst these challenging times, a glimmer of hope had emerged in the form of vaccines, offering a path to mitigate the pandemic's effects and restore normalcy. The requirement for controlled low-temperature storage during transit, however, posed a logistical nightmare, especially when it came to reaching the remote villages and underserved areas. This challenge was further compounded by the significant financial burdens it entailed for adequate control [2, 3]. Traditional vaccine transportation approaches, such as the prevalent use of insulated boxes with ice packs or phase changing material (PCM) pouches for last-mile delivery (See figure 1A and B), often resulted in temperature fluctuations and a consequential ~ 25% loss in vaccine. This loss could be attributed to vaccine vial breakage (figure 1C) and temperature fluctuation while in transport and storage, a critical concern highlighted by WHO studies [4] and GOI assessments [5]. The temperature fluctuations causing vaccine potency decline, as shown in Figure 1-E and F, arose primarily from two factors: first, the absence of proper temperature profile maintenance due to the use of phase-change materials (PCM) in insulated chambers as depicted in Figure 1C; and second, sporadic power outages in health centers located in remote regions [5]. The healthcare industry has underscored the necessity for actively monitored mini cold temperature chains (CTCs) to efficiently transport vaccines to the last mile [6]. These CTCs offer an optimal blend of cost-effectiveness and resilience, obviating the need for cumbersome transport mechanisms and prompting the redesign of existing vaccine distribution networks [7]. However, many of these cold temperature chain (CTC) systems are large and unwieldy, often unable to effectively reach the last mile where foot travel becomes necessary due to challenging terrains and limited accessibility. This is particularly prominent in countries like those in the third-world sub-Saharan region and south-east Asian countries like India.

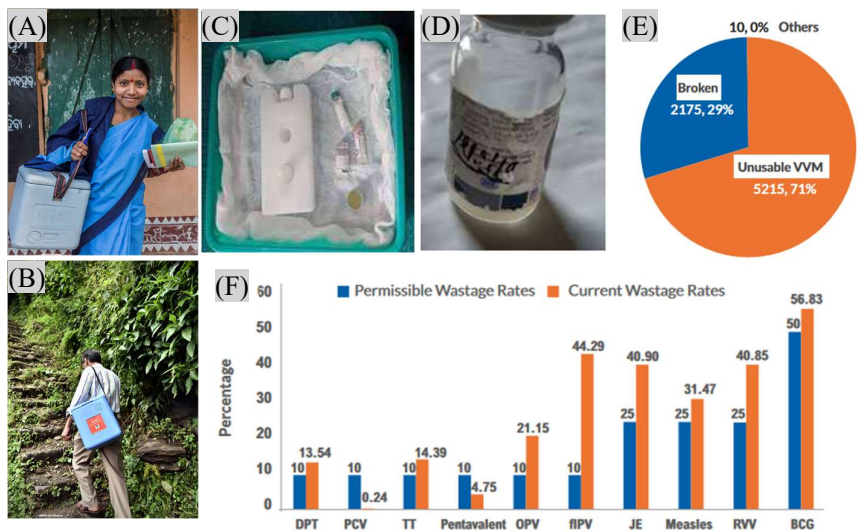

**Fig. 1** (A-B) Asha workers delivering vaccines to a remote village in southern India, using an insulated box filled with phase change material (PCM) coolant. (C) Vaccines stored inside the insulated box. (D) A vial of vaccine broken during transit. (E) Reasons for unopened vial wastage for all antigens (sample size =7,400) (F) Current vaccine wastage rates at service delivery point in comparison with permissible vaccine wastage rates. Both these reflect vaccine wastage due to inadequacies in the cold chain supply to the last-mile. The current rates are very high for most vaccines, which could be potentially reduced by the use of ALIVE module for vaccine transfer. All the pictures in this figure are adopted from [5]

Addressing this compelling need, we introduce A low-cost Logistics module with Interactive Vaccine storage Environment, abbreviated as ALIVE. Distributing vaccines to remote areas is complex due to challenges like inadequate transportation infrastructure, difficult terrains, weather conditions, maintaining active temperature control, high expenses, security risks and limited data monitoring. These hurdles lead to inefficiencies and lower vaccine quality, affecting marginalized communities. ALIVE is poised as a solution addressing complex challenges in vaccine distribution to remote areas. It's a self-sufficient off-grid storage and transport module with active cooling, ensures vaccine viability. Seamlessly integrated with IoT, ALIVE offers real-time data for informed decision-making and secure access control empowering remote monitoring and precise control over the module's internal parameters, and could be operated by even the unskilled or semi-skilled workers [8]. Its microcontroller enables precise thermophysical parameter control, minimizing wastage. ALIVE's adaptability, affordability and user-friendly design empower healthcare providers with real-time monitoring and control. This innovative technology bridges the gap between vaccine accessibility and logistical hurdles, ensuring vaccines' safe delivery to the underserved communities. Its compact, lightweight and compatibility with the state-of-the art communication technology to ensure precise temperature and humidity control, safeguarding vaccine integrity during transportation to remote regions. This study aims to showcase frugal approach through ALIVE, aiming to narrow the divide between vaccine availability and complex logistical challenges, empowering healthcare providers to deliver essential vaccines to the remotest areas and under unforgiving environmental conditions.

## 2    The ALIVE's Construct

Figure 2 outlines the operational framework of the ALIVE system. This system is structured into three integral components: the portable storage chamber (depicted in Figure 2A-i), an integrated control module (Figure 2A-ii) and an operation interface module, functioning as a control center. This module can manifest as a computer located at a central operational facility or as a mobile phone carried by personnel (represented in Figure 2A-iii). Additionally, there exists an IoT-driven data sharing platform comprising a cloud server, exemplified here by a laptop (shown in Figure 2A-iv).

The portable storage module is composed of an insulated enclosure housing a Peltier element designed for both heating and cooling functions, thereby maintaining an optimal thermophysical environment as required. Within this chamber, vaccine and medicine vials are securely stored. Notably, the module provides versatility in operation in both on-grid and off-grid scenarios. In typical conditions, it draws power from standard wall-outlets. For mobile use, the module is equipped with a battery pack to ensure uninterrupted functionality.

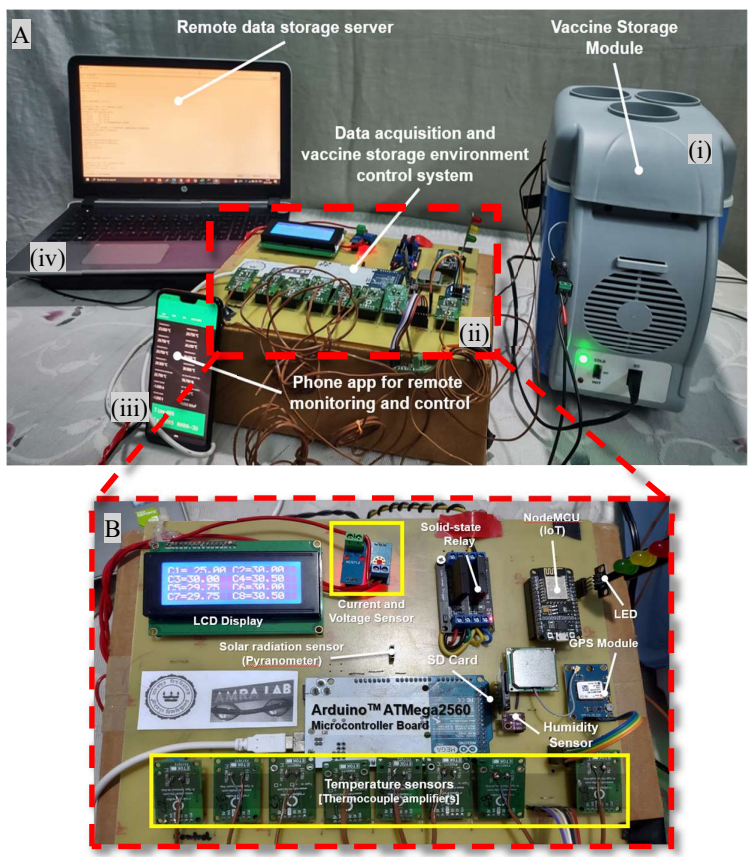

**Fig. 2** depicts the ALIVE system's three integral components: (2A-i) the portable storage chamber, (2A-ii) integrated control module, (2A-iii) IoT-driven data sharing and interactive platform and operation interface module with (2A-iv) a cloud server. The interactive platform can be a central computer or a mobile phone. The storage chamber maintains an optimal thermophysical environment with careful operation of actuators, a Peltier element in the present case. The sensor-board (2A ii and 2B) integrates essential components, including temperature and humidity sensors, GPS, and a microcontroller for precise control and data transmission.

Illustrated in Figure 2A ii, the ALIVE system incorporates a sensor-board, seamlessly integrating the pivotal components encompassing temperature and humidity sensors (SHT31), voltage and current sensors (a generic module with an embedded MAX471 chip), a GPS module (Ublox NEO-6M GPS sensor), display unit(s), LEDs and is all articulated by a singular Arduino microcontroller (Arduino™ Mega), integrated onto a unified circuit board, as revealed in 2B. This control module orchestrates the precise maintenance of thermophysical parameters within the storage chamber, by transmitting digital/analog commands, through relays (Solid State Relay Module, generic make), to actuators (viz. Peltier's, fans, chillers, and heaters). The control module is designed with this inherent adaptability to allow the end-users to customize the sensor configuration according to specific requirements. The control module offers an interactive interface (Figure 2A iii) for overseeing the storage chamber's environment, enabled by the microcontroller board's adept data processing. Captured data is stored on a micro-SD card (generic make) and seamlessly sent to a remote server via an IoT device (ESP8266 module). Accessible via a laptop, the remote server (Figure 2A iv) enables real-time monitoring, control, and data visualization. Figure 3 outlines the IoT-based data transfer process, featuring a schematic algorithm that illustrates the sequential data flow among ALIVE components. This algorithm employs a feedback loop, enabling interaction between sensors and actuators to uphold user-defined environmental conditions.

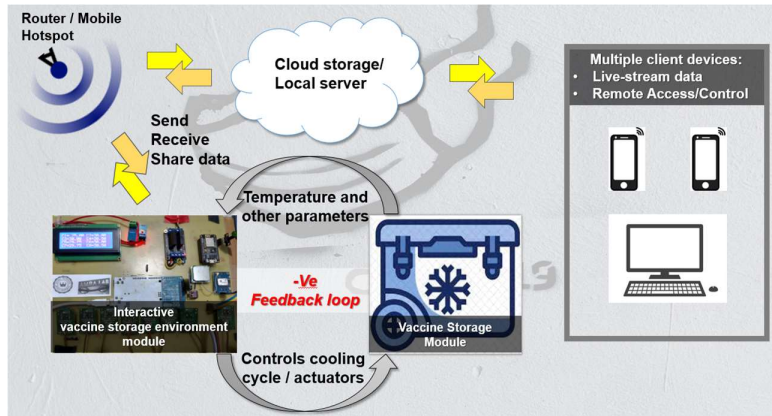

**Fig. 3** presents the data transfer process, illustrating sequential data flow among ALIVE components. The negative feedback loop enables interaction between sensors and actuators, maintaining user-defined conditions communicated via the cloud and interactive module. ALIVE harmonizes data storage, IoT communication, and parameter management for remote adjustment and precise storage conditions.

These conditions are communicated to the end user through the cloud and displayed on the interactive module (Figure 2A iii), manifested here by an API in a mobile phone. This adaptable methodology can be applied to various user-defined thermal and flow systems. ALIVE operates harmoniously by concurrently storing data, communicating with the IoT platform and managing system parameters. This empowers users to remotely adjust and optimize target operating parameters for precise storage conditions. Additionally, fine-tuning can also be performed using the onboard data controls.

In the present scenario, the Peltier element integrated into the ALIVE module for thermophysical conditioning of the storage chambers has a power rating of ~ 100W, (12V and 6A). Its operational range spans from 2°C to 98°C, facilitating a wide spectrum of temperature control. The battery capacity is determined based on anticipated operational requirements under off-grid settings, factoring in range and duration. Rather than continuous operation, there is an intermittence of the Peltier element's operation and the element is selectively powered as needed. This strategic activation is governed by the Arduino-powered control module embedded within the ALIVE system, optimizing its energy usage and ensuring precise temperature management.

## 3    Operation of the ALIVE module

The ALIVE module employs the PID (Proportional-Integral-Derivative) algorithm to achieve precise temperature control within the vaccine chamber. The PID algorithm dynamically adjusts the actuator based on the interplay of its three key components: Proportional (P), Integral (I), and Derivative (D). This synergy results in a control output that adaptively maintains the desired temperature, enabled by fine-tuning coefficients ($k_p$, $k_i$, $k_d$) specific to the system's requirements. In practice, the PID algorithm continuously evaluates the deviation between the actual and target temperature, generating the PID_value. This value determines the actuator's activation duration, ensuring the temperature remains on target. Importantly, the algorithm's coefficients are not universally applicable; they necessitate auto-tuning according to each application's nuances, facilitated by auto-tuning techniques.

To showcase ALIVE's capabilities, we aim to monitor temperature profiles in a portable refrigerated unit, pertinent for swift vaccination drives. Figure 4 shows the day-time and night-time operation modes of a typical vaccine chamber, with distinct lid usage. The algorithm compensates for temperature fluctuations caused by door openings during the day, warranting a lower temperature set-point, while the night-time set point is raised to minimize power consumption and sample-freezing. This approach maintains "safe" temperature limits for vaccine storage, allowing for energy-efficient operation without compromising safety. Experiments, as demonstrated, validate the system's ability to uphold stable temperatures despite variable conditions, endorsing its practical efficacy and potential for sustainable vaccine storage solutions. The system optimizes energy while ensuring safe vaccine storage. In a room held at ~24°C, we denote day-time temperature at 15°C and night-time temperatures at 20°C and the modes were simulated, each spanning ~3 hours within a 6-hour period. The plots in Figure 4 describe the ambient atmosphere, vaccine storage, and representative vaccine pouch temperatures. The Arduino automatically ensures the set-point temperatures, activating the Peltier element as needed. We opened and closed the module's door multiple times to simulate operation during experimentation. Despite door openings, the PIROEC maintains vaccine pouch temperatures consistently within statistical limits (SD ~ 0.6), attesting to its robust temperature control performance. The temperature set-points chosen in the study are selected arbitrarily and should be viewed as illustrative examples. However, the Arduino-controlled ALIVE module, with its Peltier element, has the capability to attain the requisite temperatures for the secure storage of a majority of Covid-19 vaccines [9], thereby safeguarding their potency.

The ALIVE IoT-enabled API offers a comprehensive array of functions. Depicted in Figure 5, within the ALIVE module's IoT-based data monitoring interface, referred to as the LIVE TICKER, takes the center stage.

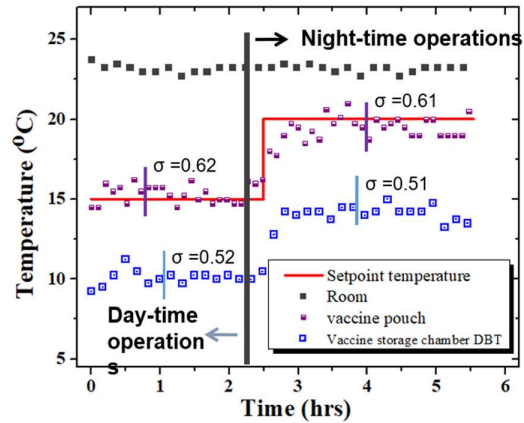

**Fig. 4** demonstrates ALIVE's prowess by monitoring temperature profiles in a portable refrigerated unit during day and night modes. The system adeptly maintains stable vaccine storage temperatures, even in the face of variable conditions, affirming its reliability and energy-efficient design.

This interface comprises three distinct tabs, each serving a unique purpose. The first tab, described in Fig. 5A is a live data streaming, which provides an overview of key environmental parameters, including temperature, humidity, and power consumption. It offers real-time insights into system conditions, displaying acquired data sets. In case of any data acquisition issues, an error message (-1) is indicated. The connection duration is also showcased. This tab can also be programmed to visually represent both real-time and historical data from various ALIVE module sensors. Graphical representations aid in easy interpretation. The second tab (Fig 5B) is the GPS Location and live monitoring tab, which presents live location data of the ALIVE module using an onboard Ublox NEO-6M GPS sensor. The third tab (Fig. 5C) offers remote access and precise control over temperature profiles within the vaccine storage module. This empowers personnel situated in a centrally located control room to actively monitor the health and viability of the vaccine through its thermo-physical parameters. Users are equipped with the ability to effortlessly fine-tune the temperature settings using the intuitive temperature change slider. Furthermore, as depicted in Figure 5C inset, the ALIVE system showcases its remote actuation by effectively toggling three LEDs ON and OFF. This exemplifies the possibility of actuating a multitude of actuators (instead of just switching of the LEDs), for example a stepper motor leading to the lid-opening and closing, or a digital lock for restricting unauthorized access.

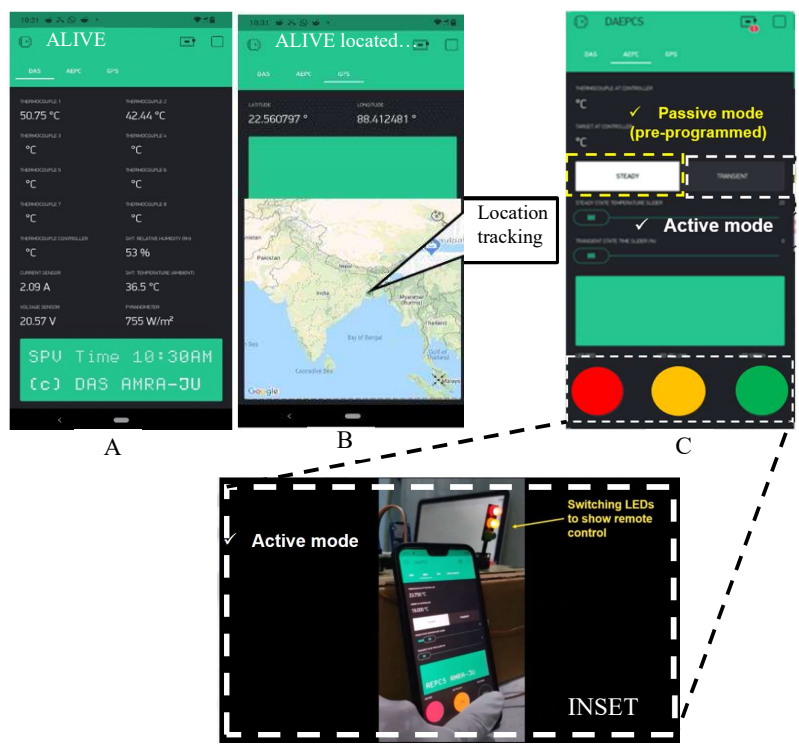

**Fig. 5** Demonstration of the LIVER-TICKER (A) The Real-time Environmental Monitoring tab dynamically streams essential metrics, including temperature, humidity, and power consumption from the ALIVE module. This interactive tab presents both real-time and historical data from various ALIVE module sensors, presented through user-friendly graphical representations, for easier interpretation. (B) The 'GPS Location and Live Monitoring' tab displays real-time geographical positioning data of the ALIVE module through its integrated onboard GPS sensor. (C) The third tab enables remote access and precise control of temperature profiles in the vaccine storage module, empowering centralized personnel to actively monitor vaccine health and adjust settings effortlessly using an intuitive slider. INSET of C: Demonstrates of remote actuation of ALIVE by efficiently toggling three LEDs over the internet using a smartphone, exemplifying versatility for creative applications and controlled access via centralized control room authorization.

The ALIVE module, developed with a frugal microcontroller foundation and integrated sensors, offers remarkable flexibility by not only measuring and storing data but also taking action based on defined algorithms. A key requirement in this regard is the necessity for sensors compatible with the Arduino platform. Envisaged as a multi-modal system, the ALIVE platform's applications range from laboratory-scale steady-state or time-variant parameter control to remote monitoring of biomedical data and medication implementation. With real-time geo-tracking capabilities, it proves valuable in last-mile vaccine logistics and environmental control.

The scalability, affordability, and adaptability of ALIVE position it for industrial integration, catering to various logistical needs and enhancing economic sustainability. The potential applications also span from FMCG logistics and livestock management to reducing material wastage, even material handling in hazardous environments. Future prospects of the ALIVE module encompass drone-assisted vaccine delivery, point-of-care diagnostics, automated medicine delivery, and micro-renewable energy control. The system's versatility extends to environment-controlled supply chains and distributed weather monitoring, showcasing its substantial potential for multifaceted applications.

## 4 Cost comparison and future direction

The creation of the ALIVE module incurred an approximate cost of INR 5000, including the development expenses (like burnt circuits and replacements) linked to the novel system creation. However, expenses related to personal devices like mobile phones, laptops, or internet services used in the development process were excluded from this calculation. With economies of scale anticipated, this cost is likely to decrease. The current price point is justified by the notable advantages offered by the device, which overshadows the costs associated with personnel training and infrastructure enhancement for ALIVE's integration. These benefits include mitigating vaccine losses and facilitating remote monitoring. While the system currently stands at Technology Readiness Level (TRL) 4, collaboration with potential healthcare partners are sought to elevate its TRL and expand its impact.

## 5    Conclusion

The ALIVE module is poised as a transformative solution to the intricate challenges posed by vaccine and medicine distribution in the face of the varied challenges of adverse topography of the land and inadequate medical infrastructure. By seamlessly integrating an established technology into a frugal architecture, ALIVE addresses the critical issues of temperature stability, logistical complexities and remote monitoring. Its design ensures optimal vaccine storage conditions, projecting a significantly reduced vaccine wastage and preserving potency. The device has been demonstrated to provide active temperature control under a simulated real-life temperature cycling of a vaccine box within statistical limits (SD ~ 0.6), and remote actuation over the internet. The modular nature of the ALIVE's extends its potential applications far beyond vaccine logistics, promising advancements in various fields. With its cost-effectiveness and scalability, ALIVE holds the promise of revolutionizing the way vaccines are transported, bridging the gap between accessibility and challenging logistics.

## Acknowledgement

AD gratefully acknowledge the funding through projects COE for Phase Transformation and product Characterization, as a part of TEQIP – III, and SERB/CRG-2019/005887, AM gratefully acknowledges the funding through RUSA 2.0 scheme of JU.

## Declaration of competing interests

AD, AM, AD and RG are the inventors on a patent application titled "A Portable, Integrated, Remotely Operable Environmental Chamber with real-time location tracking for multimodal, multi-system applications" related to this work filed with Indian patent office (application no: 202231030204, filed on 26th May, 2022).